\documentclass[twocolumn,showpacs,prb,amsmath,amssymb]{revtex4}
\usepackage{graphicx}
\usepackage{dcolumn}
\usepackage{bm}

\usepackage{makeidx}
\usepackage{epsfig}
\usepackage{subfigure}
\usepackage{wrapfig}
\input comment.sty

\begin{document}

\font \tenrm=cmr10 at 8pt

\newif \ifdraft
\drafttrue
\draftfalse

\def\half{\frac{1}{2}}

\title{Intrinsic optical bistability of thin films of linear
molecular aggregates: The two-exciton approximation}

\author{Joost A. Klugkist}
\author{Victor A. Malyshev}
\author{Jasper Knoester}
\affiliation{ Centre for Theoretical Physics and Zernike Institute
for Advanced Materials, University of Groningen, Nijenborgh 4, 9747
AG Groningen, The Netherlands }
\date{\today}

\begin{abstract}

We generalize our recent work on the optical bistability of thin
films of molecular aggregates [J. Chem. Phys. {\bf 127}, 164705
(2007)] by accounting for the optical transitions from the
one-exciton manifold to the two-exciton manifold as well as the
exciton-exciton annihilation of the two-exciton states via a
high-lying molecular vibronic term. We also include the relaxation
from the vibronic level back to both the one-exciton manifold and
the ground state. By selecting the dominant optical transitions
between the ground state, the one-exciton manifold, and the
two-exciton manifold, we reduce the problem to four levels, enabling
us to describe the nonlinear optical response of the film. The one-
and two-exciton states are obtained by diagonalizing a Frenkel
Hamiltonian with an uncorrelated on-site (diagonal) disorder. The
optical dynamics is described by means of the density matrix
equations coupled to the electromagnetic field in the film. We show
that the one-to-two exciton transitions followed by a fast
exciton-exciton annihilation promote the occurrence of bistability
and reduce the switching intensity. We provide estimates of
pertinent parameters for actual materials and conclude that the
effect can be realized.

\end{abstract}


\maketitle

\section{Introduction}

The phenomenon of optical bistability already has more than thirty
years of history, going back to the theoretical prediction of
McCall~\cite{McCall74} in 1974, followed by experimental
demonstration of the effect by Gibbs, McCall, and
Venkatesan~\cite{Gibbs76} in 1976 (see
also~Refs.~\onlinecite{Lugiato84,Gibbs85},
and~\onlinecite{Rosanov96} for an overview). Since then a vast
amount of literature has been devoted to explore the topic (an
extended bibliography can be found in our recent paper,
Ref.~\onlinecite{Klugkist07a}); controlling the flow of light by
light itself is of great importance for optical technologies,
especially on the micro- and nano-scale. More recently, new
materials such as photonic crystals,~\cite{Soljacic03}
surface-plasmon polaritonic crystals,~\cite{Wurtz06} and materials
with a negative index of refraction,~\cite{Litchinitser07} have
revealed bistable behavior.

In our previous work,~\cite{Klugkist07a} we studied theoretically
the bistable optical response of a thin film of linear molecular
J-aggregates. To describe the optical response of a single
aggregate, we exploited a Frenkel exciton model with an uncorrelated
on-site energy disorder, taking into account only the optically
dominant transitions from the ground state to the one-exciton
manifold, while neglecting the one-to-two exciton transitions.
Within this picture, an aggregate can be viewed as a meso-ensemble
of two-level localization segments,~\cite{Malyshev00} which allows
for a description of the optical dynamics by means of a $2\times
2$-density matrix. Employing a joint probability distribution of the
transition energy and the transition dipole moment of Frenkel
excitons, allowed us to account for the correlated fluctuations of
these two quantities, obtained from diagonalizing the Frenkel
Hamiltonian with disorder. By solving the coupled Maxwell-Bloch
equations, we calculated the phase diagram of possible stationary
states of the film (stable, bistable) and the input-dependent
switching time. From the analysis of the spectral distribution of
the exciton population at the switching point, we realized that the
field inside the film is sufficient to produce one-to-two exciton
transitions, confirming a similar statement raised in
Ref.~\onlinecite{Glaeske01}.

%Fig. 1

\begin{figure}[lht]
\begin{center}
\includegraphics[width = .45\textwidth,scale=1]{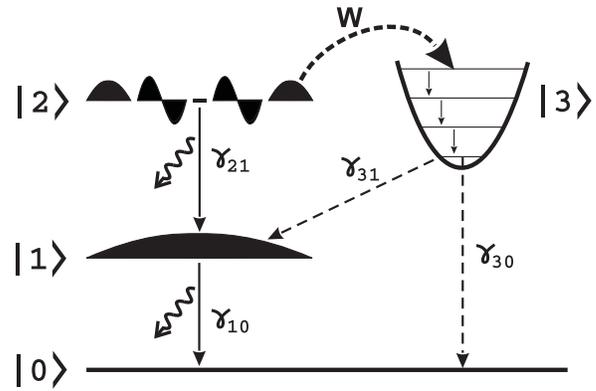}
\end{center}
\caption{Four-level model of the film's optical response.
    The input field induces transitions between the ground state $|0 \rangle$,
    one-exciton manifold $|1\rangle$, and two-exciton manifold $|2\rangle$.
    The population of the latter is transferred with a rate $w$ to a
    vibronic molecular level $|3\rangle$, followed by
    fast relaxation between the vibronic sublevels towards the
    vibronic ground state. Finally, the latter undergoes relaxation
    to the one-exciton and/or ground state with the rates $\gamma_{31}$
    and $\gamma_{30}$, respectively. The constants $\gamma_{10}$ and
    $\gamma_{21}$ denote the radiative decay of the one- and
    two-exciton states, respectively.}
    \label{Fig: Four level model}
\end{figure}

The goal of the present paper is to extend the one-exciton
model~\cite{Klugkist07a} by including two-exciton states and
transitions between the one- and two-exciton manifolds,
respectively. Furthermore, two excitons spatially located within the
same localization domain usually quickly annihilate, transferring
their energy to an appropriate resonant monomer vibronic
level.~\cite{Stiel88,Sundstroem88,Minoshima94,Gaizauskas95,Scheblykin00a,Shimizu01,Brueggemann03,Spitz06}
Hence, the generalized model requires the consideration of
exciton-exciton annihilation.~\cite{Glaeske01} We will assume that
exciton-exciton annihilation prevents the three-exciton states from
playing a significant role in the response of the film. The relevant
transitions of the model are depicted in Fig.~\ref{Fig: Four level
model}.

To make the two-exciton model tractable, we will select the
optically dominant transitions between the ground state and the
one-exciton manifold (as we did in~Ref.~\onlinecite{Klugkist07b}),
and also between the one- and two-exciton manifolds. Treating the
different localization segments independently, in combination with
the state selection, allows one to considerably reduce the set of
relevant states, namely to four states of a segment: the ground
state, the optically dominant one- and two-exciton states, and a
high-lying molecular electronic or vibronic state, through which the
excitons annihilate. This model has been implemented for the first
time in Ref.~\onlinecite{Glaeske01}, using the simplifying
assumption that the transition energies and transition dipole
moments are correlated perfectly. Unlike
Ref.~\onlinecite{Glaeske01}, we will account for the correct joint
statistics of both quantities, similarly to our previous
work.~\cite{Klugkist07a} The optical dynamics of a single
localization segment is described within the framework of a $4\times
4$-density matrix. We derive a steady-state equation for the output
field intensity as a function of the input intensity. The field
inside the film is calculated taking into account the field produced
by the aggregate dipoles. We find that, counterintuitively, tuning
away from the resonance may, depending on the dephasing rate,
promote bistable behavior. In addition, we show that fast
exciton-exciton annihilation combined with slow relaxation from the
high-lying vibronic level enhances the tendency towards bistablity.
The phase diagram of bistability is computed and compared with the
one-exciton model. We address also the realizability of the bistable
behavior in actual materials.

The outline of this paper is as follows. In the next section we
present our model of a single aggregate, consisting of a Frenkel
Hamiltonian with uncorrelated on-site energy (diagonal) disorder
(Sec.~\ref{Sec: Single agregate}). Next, we describe the selection
of the optically dominant transitions in Sec.~\ref{Sec: Selection
procedure} and introduce our model for the exciton-exciton
annihilation of two-exciton states in Sec.~\ref{Sec: Ex-ex
annihilation}. In Sec.~\ref{Sec: Density matrix-field equations}, we
formulate our approach, based on the density matrix equations in the
$4\times4$ space of states, as well as the Maxwell equation for a
thin film of oriented linear J-aggregates. Section~\ref{Sec:
Steady-state analysis} deals with the results of our numerical
analysis of the bistable optical response of the film in a
multidimensional parameter space. We identify conditions that are
most favorable for bistable behavior of the film. In Sec.~\ref{Sec:
Estimates}, we estimate the driving parameters and the input light
flux required for experimental realization of bistability for films
of pseudoisocyanine J-aggregates. Section~\ref{Summary} summarizes
the paper.

\section{Model}
    \label{Sec: Model}

The geometry of the model system and the assumptions we adopt
hereafter are essentially the same as in our previous
paper.~\cite{Klugkist07a} In short, we aim to study the
transmittivity of an assembly of linear J-aggregates arranged in a
thin film (with the film thickness $L$ small compared to the
emission wavelength $\lambda^{\prime}$ inside the film) and aligned
in one direction, parallel to the film plane. The aggregates in the
film are assumed to be decoupled from each other; their coupling to
the environment is treated through phenomenological relaxation rates
(see Ref.~\onlinecite{Klugkist07a} for a detailed discussion).

\subsection{A single aggregate}
    \label{Sec: Single agregate}

We model a single aggregate as a linear chain of $N$ three-level
monomers. The two lower states are assumed to form multi-exciton
bands, as a result of strong dipole-dipole excitation transfer
interactions between the monomers. To simplify the treatment of the
multi-exciton states, we restrict ourselves to nearest-neighbor
interactions. The transition dipole moments between the two lower
molecular states are considered to align in one direction for all
monomers. Then the (Frenkel) exciton part of the aggregate
Hamiltonian reads
\begin{equation}
    H_0 = \sum_{n=1}^N \> \epsilon_n b_n^{\dagger}b_n  -
    J \sum_{n=1}^{N-1} \left( b_n^{\dagger} b_{n+1}
    + b_{n+1}^{\dagger} b_{n}\right ) \ ,
\label{H}
\end{equation}
where $b_n^{\dagger}(b_n)$ denotes the creation (annihilation) Pauli
operator of an excitation at site $n$. The monomer excitation
energies $\epsilon_n$ between the two lower states are modeled as
uncorrelated Gaussian variables with mean $\epsilon_0$ and standard
deviation $\sigma$. The parameter $J$ represents the magnitude of
the nearest-neighbor transfer integral. We assume that it does not
fluctuate. After applying the Jordan-Wigner transformation, the
multi-exciton eigenstates are found as Slater determinants of
one-exciton states $\varphi_{\nu n}$ with different
$\nu$.~\cite{Chesnut63,Juzeliunas88,Spano91} The multi-exciton
eigenenergies are given by $\sum_{\nu=1}^N
n_{\nu}\varepsilon_{\nu}$, with $\varepsilon_{\nu}$ being the
one-exciton eigenenenrgies and $n_{\nu}= 0,1$ depending on whether
the $\nu$th state is occupied or not. Particularly, we will be
interested in the one- and two-exciton states:
\begin{subequations}
    \label{one- and two-exciton states}
\begin{equation}
    \label{one-exciton states}
    |\nu \rangle = \sum_{n=1}^N \varphi_{\nu n} |n \rangle \ ,
\end{equation}
\begin{equation}
    \label{two-exciton states}
    |\nu_1 \nu_2 \rangle
    = \sum_{n_1 > n_2}^N \left(\varphi_{\nu_1 n_1}\varphi_{\nu_2 n_2}
    - \varphi_{\nu_1 n_2}\varphi_{\nu_2 n_1} \right) |n_1 n_2
    \rangle \ ,
\end{equation}
\end{subequations}
where $|n \rangle = b_n^{\dagger} |0 \rangle$ and $|n_1 n_2 \rangle
= b_{n_1}^{\dagger}b_{n_2}^{\dagger} |0 \rangle$, and $|0 \rangle$
is the ground state of the aggregate (with all monomers in the
ground state). We also will need the transition dipole moments from
the ground state $|0 \rangle$ to a one-exciton state $|\nu \rangle$
and from a one-exciton state $|\nu \rangle$ to a two-exciton state
$|\nu_1\nu_2 \rangle$. In units of the single-molecule transition
dipole moment, they obtain the dimensionless form
\begin{subequations}
    \label{Eq: oscillator strengths}
\begin{equation}
    \label{Eq: ground-to-one}
    \mu_{\nu} = \sum_{n=1}^N \varphi_{\nu n} \ ,
\end{equation}
\begin{align}
    \label{Eq: one-to-two}
    \mu_{\nu_1\nu_2,\nu} & =
    \sum_{n_2>n_1} \left(\varphi_{\nu n_1} - \varphi_{\nu n_2} \right)
    \nonumber\\
    \nonumber\\
    & \times
    \left(\varphi_{\nu_1 n_1}\varphi_{\nu_2 n_2}
    - \varphi_{\nu_1 n_2}\varphi_{\nu_2 n_1} \right) \ ,
\end{align}
\end{subequations}
where it was assumed that the aggregate is small compared to an
optical wavelength.

\subsection{Selecting the dominant exciton transitions}
    \label{Sec: Selection procedure}

At low temperatures, exciton states reduce their extension from the
physical size of the aggregate to much smaller segments as a result
of the disorder-induced Anderson
localization~\cite{Abrahams79,Schreiber82} . We will denote the
typical size of these segments as $N^*$, often referred to as the
number of coherently bound molecules or localization length in terms
of the localization theory.

For J-aggregates, the optically dominant localized states reside in
the neighborhood of the bottom of the exciton band. Some of them
resemble $s$-like atomic states: they consist of mainly one peak
with no node within the localization segment (see Fig.~\ref{Fig:
Reduced wavefunctions}{\it a}). We will denote the subset of such
states as ${\cal S}$. To find all the $s$-like states from the
complete set of wave functions $\varphi_{\nu n}$, we used the rule
proposed in Ref.~\onlinecite{Malyshev01}, $\big|\sum_n \varphi_{\nu
n} |\varphi_{\nu n}|\big|  \ge C_0$ with $C_0 = 0.75$. The
inequality selects those states that contain approximately 75\% of
the density in the main peak. We found numerically that for a wide
range of the disorder strength $\sigma$ ($0.05 J < \sigma < J$), the
thus selected states accumulate on average 73\% of the total
oscillator strength (equal to $N$). Recall that for a disorder-free
aggregate, the optically dominant (lowest) exciton state contains
81\% of the total oscillator strength of the one-exciton transitions
(see, e.g., Refs.~\onlinecite{Fidder91}
and~\onlinecite{Malyshev91}). Furthermore, we have shown that the
$s$-like states, used as a basis to calculate the linear absorption
spectrum, well reproduce its peak position and the shape of its red
part, failing slightly in describing the blue wing, where
higher-energy exciton states contribute as well.~\cite{Klugkist07b}
From this, we conclude that our procedure to select the optically
dominant ($s$-like) one-exciton states works well.

%Fig. 2

\begin{figure}[lht]
\begin{center}
\includegraphics[width = .5\textwidth,scale=1]{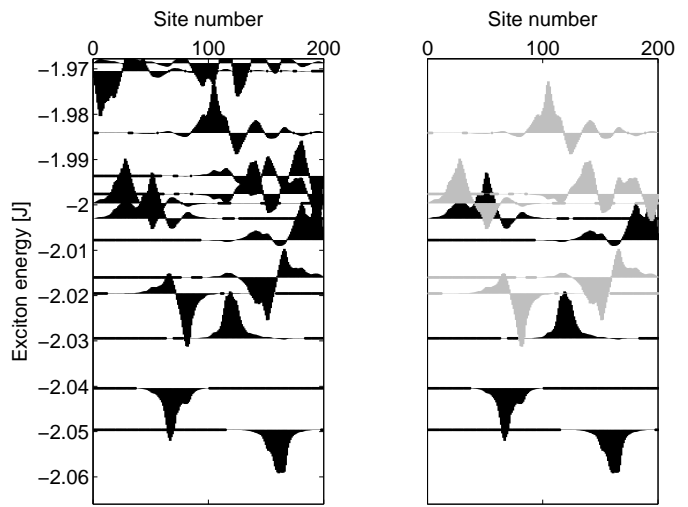}
\end{center}
\caption{(a)~The lowest 12 one-exciton states of a chain of
    length $N=500$ for a particular disorder realization at the disorder
    strength $\sigma = 0.1 J$. (b)~A subset of $s$ states (black) and
    $p_s$ states (gray) that mostly contribute to the ground state to
    one-exciton and to the one-to-two exciton transitions. The average
    single molecule transition energy $\epsilon_0$ was chosen as origin
    of the energy scale.}
    \label{Fig: Reduced wavefunctions}
\end{figure}

Similar to the $s$-like states, one may also distinguish states that
resemble atomic $p$ states. They have a well defined node within
localization segments and occur in pairs with $s$-like states. Each
pair forms an $sp$ doublet localized on the same chain segment. The
levels within a doublet undergo quantum level repulsion, with their
spacing nicely following the one that exists between $k=1$ and $k=2$
exciton states in a homogeneous chain of size
$N^*$.~\cite{Malyshev01} From the theory of multi-exciton
transitions in homogeneous aggregates,~\cite{Knoester93} we know
that the Slater determinant of the $k=1$ and $k=2$ states forms the
two-exciton state that predominantly contributes to the two-exciton
optical response. This gives us a solid ground to believe that the
$s$-like one-exciton states and the two-exciton states composed of
$(sp)$ doublets dominate the one-to-two exciton transitions in
disordered aggregates (see below).

Usually, well defined $(sp)$ doublets occur below the bare exciton
band edge at the energy $-2J$. These doublets are responsible for a
hidden level structure of the Lifshits tail.~\cite{Malyshev95} For
the $s$-like states located close to or above the bare band edge, it
is already impossible to assign a $p$-like partner localized on the
same segment: higher-energy states have more than one node and
spread over segments of size larger than $N^*$ (see Fig.~\ref{Fig:
Reduced wavefunctions}). To obtain all the states that give a major
contribution to the one-to-two exciton transitions, the following
procedure has been used. First, we selected all the $s$-like states,
as described above. After that, we considered all the two-exciton
states $|s\nu \rangle$ given by Eq.~(\ref{two-exciton states}) and
calculated the corresponding transition dipole moments
$\mu_{s\nu,s}$. From the whole set of $\mu_{s\nu,s}$, we took the
largest ones denoted by $\mu_{sp_s,s}$, were the substrict $s$ in
$p_s$ indicates its relation with the state $|s\rangle$. This
procedure catches all true $sp_s$ doublets and assigns a partner to
solitary $s$-like states, which do not necessarily look like real
$p$ states. In Fig.~\ref{Fig: Reduced wavefunctions}{\it b}, we
depicted the final set of the doublets selected from the states in
Fig.~\ref{Fig: Reduced wavefunctions}{\it a} according to the above
procedure, which contribute mostly to the one- and two-exciton
transitions.

The average ratio of the oscillator strength of the thus selected
transitions $|s\rangle\to |sp_s\rangle$ and $|0\rangle\to |s\rangle$
turned out to be approximately 1.4. For a homogeneous chain, this
ratio equals 1.57 (then $|s \rangle = |k=1 \rangle$ and $|sp_s
\rangle = |k_1=1,k_2=2 \rangle$). The similarity of these numbers
gives support to our selection procedure. Even stronger support is
obtained from computing the pump-probe spectrum, using our state
selection, and comparing the result to that of the exact
calculations.~\cite{Klugkist07b} The comparison revealed that the
model spectrum only deviates from the exact one in the blue wing of
the induced absorption peak, similarly to the linear absorption
spectra.

\subsection{Exciton-exciton annihilation}
    \label{Sec: Ex-ex annihilation}

As was already mentioned in the Introduction, two excitons created
within the same localization segment efficiently annihilate (the
intra-segment annihilation in terms of Refs.~\onlinecite{Malyshev99}
and~\onlinecite{Ryzhov01}). Thus, the authors of
Ref.~\onlinecite{Minoshima94} studied experimentally the exciton
dynamics in J-aggregates of pseudoisocyanine bromide (PIC-Br) at low
temperature and found a 200 fs component in the two-exciton state
decay. They attributed this to the annihilation of two-excitons
located within the same chain segment of typical size of $N^* = 20$.
We adopt this mechanism for $|sp_s \rangle$ states described in the
preceding section. Note that 200 fs is much shorter than all other
population decay times. Other processes, such as radiative decay,
occur at times of tens-to-hundreds of picoseconds.

Two excitons located on different localization segments can also
annihilate (the inter-segment annihilation in terms of
Refs.~\onlinecite{Ryzhov01}). This process, however, is much slower
as compared to the intra-segment channel;~\cite{Ryzhov01} we neglect
it. The thermally activated diffusion of excitons accelerates the
annihilation of excitons created far away from each other. We
consider this diffusion-limited exciton annihilation as irrelevant
to our problem, because for bistability to occur we need the
majority of $s$-like states to be saturated (also see Sec.~\ref{Sec:
Estimates}).

It is usually assumed that the annihilation occurs via transferring
the two-exciton energy to a resonant molecular vibronic level (see,
e.g., Ref.~\onlinecite{Stiel88}), which undergoes a fast
vibration-assisted relaxation to the ground vibronic state. The
population collected in this state relaxes further to the
one-exciton state $|1\rangle$ of the segment or to the ground state
$|0\rangle$ of the aggregate (cf. Fig~\ref{Fig: Four level model}).
In this way, one or two excitations, respectively, are taken from
the system. In summary, a four-level model, including the ground,
one- and two-exciton states, and a molecular vibronic level through
which the excitons annihilate, should be employed to describe the
optical response of the film in the two-exciton approximation.

\subsection{Truncated density-matrix-field equations}
    \label{Sec: Density matrix-field equations}

Within the four-level model introduced in the preceding sections, we
describe the optical dynamics of a segment in terms of a $4\times 4$
density matrix $\rho_{\alpha\beta}$, where the indexes $\alpha$ and
$\beta$ run from 0 to 3, where $|1 \rangle \equiv |s \rangle$ and
$|2 \rangle \equiv |sp_s \rangle$. We neglect the off-diagonal
matrix elements $\rho_{30}, \rho_{31}$, and $\rho_{32}$, assuming a
fast vibronic relaxation within the molecular level 3. Within the
rotating wave approximation, the set of equations for the
populations $\rho_{\alpha\alpha}$ and for the amplitudes of the
relevant off-diagonal density matrix elements $R_{\alpha\beta},
(\alpha \ne \beta)$ reads~\cite{Glaeske01}
\begin{widetext}
\begin{subequations}
    \label{Eq: Density matrix truncated}
\begin{equation}
    \dot{\rho}_{00} = \frac{1}{4} \mu_{10} \left[\Omega R^*_{10}
    + \Omega^* R_{10} \right]+ \gamma _{10}\rho_{11} + \gamma _{30}\rho_{33}
\ ,
\end{equation}
\begin{equation}
\label{Eq: rho11}
    \dot{\rho}_{11} = -\gamma_{10} \rho_{11}+ \gamma_{21} \rho_{22}
    + \gamma_{31} \rho_{33} + \frac{1}{4}  \mu_{21} \left(\Omega R^*_{21}
    - \Omega^*R_{21} \right)
   -\frac{1}{4} \mu_{10} \left(\Omega R^*_{10} - \Omega^*R_{10} \right)  \ ,
\end{equation}
\begin{equation}
    \dot{\rho}_{22} = - \left( \gamma _{21} + w\right)\rho_{22}
    -\frac{1}{4} \mu_{21} \left(\Omega R_{21}^* - \Omega^* R_{21} \right) \ ,
\end{equation}
\begin{equation}
    \dot{\rho}_{33} = -\gamma_3\rho_{33}  +w\rho_{22} \ ,
\end{equation}
\begin{equation}
\label{Eq: R10}
    \dot{R}_{10} = -\left(i \Delta_{10} + \Gamma_{10}\right)R_{10}
    -\mu_{10}\Omega\left(\rho_{00} -\rho_{11} \right)
    + \frac{1}{2}i\mu_{21}\Omega^* R_{20} \ ,
\end{equation}
\begin{equation}
    \dot{R}_{21} = -\left(i \Delta_{21} + \Gamma_{21}
    + \frac{1}{2} w\right)R_{21} -\mu_{21}\Omega\left(\rho_{11}
    -\rho_{22} \right) - \frac{1}{2}i \mu_{10}\Omega^* R_{20} \ ,
\end{equation}
\begin{equation}
    \dot{R}_{20} = -\left(i \Delta_{10} +i \Delta_{21}+ \Gamma_{20}
    + \frac{1}{2} w\right)R_{20} + \frac{1}{2}i \mu_{21}\Omega R_{10}
    - \frac{1}{2} i \mu_{10}\Omega R_{21} \ .
\end{equation}
\end{subequations}
\end{widetext}
Here, $\gamma_{10} = \gamma_0 |\mu_{10}|^2$ and $\gamma_{21} =
\gamma_0 |\mu_{21}|^2$ are the radiative relaxation rates of the
one-exciton state $|1 \rangle$ and the two-exciton state $|2
\rangle$, respectively, with $\gamma_0$ denoting the monomer
radiative rate, and $\mu_{10}$ and $\mu_{21}$ being the
corresponding dimensionless transition dipole moments. Furthermore,
$w$ is the annihilation constant of the two-exciton state $|2
\rangle$ and $\gamma_{3} = \gamma_{30} + \gamma_{31} $ is the
population relaxation rate of the vibronic state $|3 \rangle$. The
constants $\Gamma_{10} = \gamma_{10}/2 + \Gamma$ and $\Gamma_{21} =
\Gamma_{20} = \gamma_{21}/2 + \Gamma$ stand for the dephasing rates
of the corresponding transitions. They include a contribution from
the population decay as well as a pure dephasing part $\Gamma$,
which, for the sake of simplicity, we assume equal for all
off-diagonal density matrix elements and not fluctuating. By
$\Delta_{10} = \omega_{10}- \omega_i$ and $\Delta_{21} =
\omega_{21}- \omega_i$ we denote the detuning between the exciton
transition frequencies $\omega_{10}$ and $\omega_{21}$, and the
frequency $\omega_i$ of the incoming field. It is worth to notice
that Eqs.~(\ref{Eq: Density matrix truncated}) automatically
conserve the sum of level populations: $\rho_{00} + \rho_{11} +
\rho_{22} + \rho_{33} = 1$.

The quantity $\Omega = d_{0}E/\hbar$ in Eqs.~(\ref{Eq: Density
matrix truncated}) is the amplitude $E$ of the field inside the film
in frequency units, where $d_0$ is the transition dipole moment of a
monomer and $\hbar$ is the Planck constant. It obeys the following
equation~\cite{Glaeske01}
\begin{equation}
    \label{Eq: field equation}
    \Omega = \Omega_i  + \Gamma_R \> \frac{N_s}{N}\,
    \Big\langle \mu_{10} R_{10} + \mu_{21}R_{21} \Big\rangle \ ,
\end{equation}
where $\Omega_i = d_{0}E_i/\hbar$ is the amplitude $E_i$ of the
incoming field in frequency units, $N_s$ is the average number of
$s$-like states in an aggregate, and $\Gamma_R = 2\pi n_0 {d_0}^2 k
L/\hbar$ is the superradiant constant, an important parameter of the
model.~\cite{Malyshev99,Glaeske01,Klugkist07a} In this expression,
$n_0$ is the number density of monomers in the film, $k$ is the
field wave number, and $L$ is the film thickness. The angular
brackets in Eq.~(\ref{Eq: field equation}) denote the average over
disorder realizations.

The set of equations~(\ref{Eq: Density matrix truncated}) forms the
basis of our analysis of the effects of one-to-two exciton
transitions, exciton-exciton annihilation from the two-exciton
state, and relaxation of the annihilation level back to the
one-exciton and ground states on the optical bistable response from
an ultrathin film of J-aggregates.
In the remainder of this paper, we will be interested in the
dependence of the transmitted field intensity $|\Omega|^2$ on the
input field intensity $|\Omega_i|^2$, following from Eqs.~(\ref{Eq:
Density matrix truncated}) and~(\ref{Eq: field equation}).

\section{Steady-state analysis}
    \label{Sec: Steady-state analysis}

\subsection{Bistability equation}
    \label{Sec: Bistability equation}

To study the stationary states of the system, we first consider the
steady-state regime of the film's optical response and set the time
derivatives in Eqs.~(\ref{Eq: Density matrix truncated}) to zero.
Furthermore, we will mostly focus on the limit of fast
exciton-exciton annihilation, assuming the annihilation constant $w$
to be largest of all relaxation constants and also much larger than
the magnitude of the field inside the film, $|\Omega|$. The reason
for the latter assumption is based on the fact that below the
switching threshold, the field magnitude $|\Omega|^2 \sim
(\gamma_0\sigma^*)$,~\cite{Klugkist07a} where $\gamma_0$ and
$\sigma^*$ are the radiative decay rate of a monomer and the half
width at half maximum (HWHM) of the linear absorption spectrum,
respectively. As $\gamma_0 \ll \sigma^*$, the magnitude of the field
is also much smaller than $\sigma^*$. Above the switching threshold,
$|\Omega|$ becomes comparable to $\sigma^*$.~\cite{Klugkist07a} The
typical HWHM of J-aggregates of PIC at low temperatures is on the
order of a few tens of cm$^{-1}$, which in time units corresponds to
one picosecond. On the other hand, the time scale of exciton-exciton
annihilation is 200 femtoseconds (see Sec.~\ref{Sec: Ex-ex
annihilation}). This justifies our assumption $|\Omega| \ll w$ and
allows us to neglect $R_{20}$ in steady-state Eqs.~(\ref{Eq: Density
matrix truncated}), because $|R_{20}|\sim |\Omega/(i\Delta_{21} +
\Gamma_{20} +w/2)|$. Within this approximation, we are able to
derive a closed steady-state equation for the $\Omega$-vs-$\Omega_i$
dependence, which reads
\begin{widetext}
\begin{align}
    \label{Eq: Steady state output field}
   |\Omega_i|^2
   = \Bigg\{\left[1 + \gamma_R \frac{N_s}{N} \bigg\langle \mu_{10}^2
   \frac{\Gamma_{10}}{\Gamma_{10}^2 + \Delta_{10}^2} \left(\rho_{00}
   - \rho_{11} \right)
   + \mu_{21}^2 \frac{\Gamma_{21}+w/2}{\left(\Gamma_{21} + w/2\right)^2
   + \Delta_{21}^2} \left(\rho_{11}-\rho_{22}\right)\bigg\rangle \right]^2
   \nonumber\\
   \nonumber\\
   + \left[ \gamma_R \frac{N_s}{N}
   \bigg\langle \mu_{10}^2 \frac{\Delta_{10}}{\Gamma_{10}^2
   + \Delta_{10}^2} \left(\rho_{00} - \rho_{11} \right)
   + \mu_{21}^2 \frac{\Delta_{21} + w/2}{\left(\Gamma_{21}
   + w/2\right)^2 + \Delta_{21}^2} \left(\rho_{11}
   - \rho_{22}\right) \bigg \rangle \right]^2 \Bigg\} |\Omega|^2 \ .
\end{align}
\end{widetext}
The steady-state populations are given by~\cite{Glaeske01}
\begin{widetext}
\begin{subequations}
\begin{equation}
    \rho_{00}-\rho_{11} =  \frac{1 + \left(1
    + w \gamma_{03}/\gamma_{10}\gamma_3 \right)S_{21}}{ 1 +  2 S_{10}
    + \left(1 + w \gamma_{03}/\gamma_{10}\gamma_3 \right)S_{21}
    + \left(3 + w /\gamma_3\right)S_{10}S_{21} } \ ,
\end{equation}
\begin{equation}
    \rho_{11}-\rho_{22} =  \frac{S_{10}}{ 1+  2 S_{10} + \left(1
    + w \gamma_{03}/\gamma_{10}\gamma_3 \right)S_{21} + \left(3
    + w /\gamma_3\right)S_{10}S_{21} } \ ,
\end{equation}
\end{subequations}
\end{widetext}
where
\begin{subequations}
\begin{equation}
    S_{10}  =  \frac{ \mu_{10}^2 |\Omega|^2}{2\gamma_{10}}
    \frac{\Gamma_{10}}{\Delta_{10}^2+\Gamma_{10}^2} \ ,
\end{equation}
\begin{equation}
    S_{21}  =  \frac{ \mu_{21}^2 |\Omega|^2}{2\left(\gamma_{21}
    + w\right)} \frac{\Gamma_{21}+w/2}{\Delta_{21}^2+\left(\Gamma_{21}
    + w/2\right)^2} \ .
\end{equation}
\end{subequations}

The terms proportional to $\mu_{21}^2$ in Eq.~(\ref{Eq: Steady state
output field}) describe the effects of the two-exciton state,
exciton-exciton annihilation, and relaxation from the vibronic level
back to the one-exciton and ground states. Equation~(\ref{Eq: Steady
state output field}) reduces to the one-exciton model considered in
our previous paper~\cite{Klugkist07a} by setting $\mu_{21} = 0$.
Similarly to the one exciton model, Eq.~(\ref{Eq: Steady state
output field}) contains a small factor $N_s/N$, absent in the
earlier paper, Ref.~\onlinecite{Glaeske01}. This smallness, however,
is compensated by the $N_s$-scaling of the average in Eq.~(\ref{Eq:
Steady state output field}): it is proportional to
$\langle(\mu_{10}^2 + \mu_{21}^2)\rangle /N_s \approx 2N/N_s \gg
1$.~\cite{Klugkist07a} Thus, the actual numerical factor in
Eq.~(\ref{Eq: Steady state output field}) is approximately 2. We
stress that, unlike previous work,~\cite{Glaeske01} Eq.~(\ref{Eq:
Steady state output field}) properly accounts for the joint
statistics of all transition energies and transition dipole moments.

It is worth to notice that the second term in the first square
brackets in Eq.~(\ref{Eq: Steady state output field}) represents the
imaginary part of the nonlinear susceptibility, while the one in the
second square brackets is its real part. Hence, we will will refer
to these terms as to absorptive and dispersive, respectively,
following the convention adapted in the standard theory of
bistability of two-level systems in a cavity.~\cite{Gibbs76}

We numerically solved Eq.~(\ref{Eq: Steady state output field}),
looking for a range of parameters
($\Gamma_R,\sigma^*,\Gamma,\gamma_{31},\gamma_{30}$) where the
output-input dependence becomes S-shaped, the precursor for
bistability to occur. In all simulations, we used linear chains of
$N = 500$ sites and the radiative constant of a monomer $\gamma_0 =
2\times 10^{-5}J$ (typical for monomers of polymethine dyes). The
exciton-exciton annihilation rate was set to $w = 5000 \gamma_0$,
corresponding to an annihilation time of 200 fs.~\cite{Minoshima94}
The average single molecule transition energy $\epsilon_0$ was
chosen as origin of the energy scale. 10000 localization segments
were considered in disorder averaging.

\begin{figure}[lht]
\begin{center}
\includegraphics[width = 0.4\textwidth,scale=1] {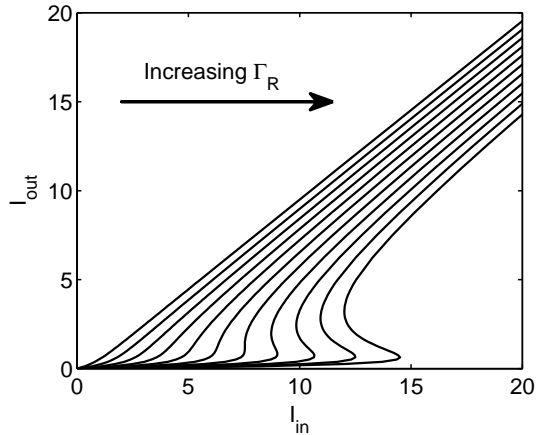}
\end{center}
\caption{Examples of the output-input characteristics,
    demonstrating the occurrence of S-shaped behavior in the film's
    optical response. Simulations were performed for a disorder strength
    $\sigma = 0.1 J$, resulting in an inhomogeneous HWHM $\sigma^* =
    0.024 J$. The incoming field was tuned to the $J$-band maximum,
    $\Delta_{10}^{(0)} = \epsilon_0 - \omega_i - 2.02 J$.
    The population relaxation rates of the vibronic state $|3 \rangle$
    were taken equal to the monomer decay rate, i.e., $\gamma_{31}=
    \gamma_{30} = \gamma_0$, while the dephasing constant $\Gamma = 500
    \gamma_0$.
    In the plot, the superradiant constant $\Gamma_R$ ranges
    from $\sigma^*$ to $11\sigma^*$ in steps of $\sigma^*$ (left to
    right). The critical value for bistability to occur is seen to be
    $\Gamma_R^c = 7\sigma^*$.}
    \label{Fig: Steady State GammaR}
\end{figure}

Figure~\ref{Fig: Steady State GammaR} shows the output intensity
$I_{\mathrm{out}} = |\Omega|^2/(\gamma_0\sigma^*)$ versus the input
intensity $I_{\mathrm{in}} = |\Omega_i|^2/(\gamma_0\sigma^*)$,
varying the superradiant constant $\Gamma_R$ from small to large
values to find the threshold for $\Gamma_R$ at which bistablity sets
in. The relaxation constants $\gamma_{30}$ and $\gamma_{31}$ from
the state $|3 \rangle$ were taken to be equal to the radiative rate
of a monomer, $\gamma_0$, which is the smallest one in the problem
under study. The incoming field was tuned to the absorption maximum
$\Delta_{10}^{(0)} = \epsilon_0 - \omega_i - 2.02 J$, which naively
speaking is expected to give the lowest threshold for bibtability
(see a discussion of the detuning effects in Sec.~\ref{Sec: Effect
of detuning}). The other parameters of the simulations are specified
in the figure caption. As follows from Fig.~\ref{Fig: Steady State
GammaR}, for the given set of parameters the bistability threshold
is $\Gamma_R^c = 7 \sigma^*$.

%Fig. 3

\subsection{Effects of relaxation from the vibronic level}
    \label{Sec: Effect of Gammas}

\begin{figure}[lht]
\begin{center}
\includegraphics[width = 0.9\columnwidth,scale=1]{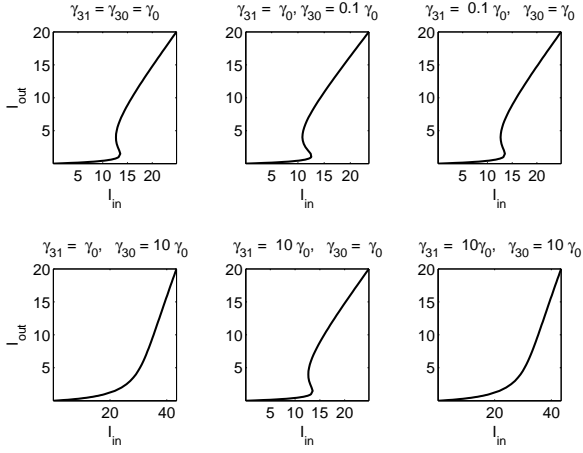}
\end{center}
    \caption{Examples of the output-input characteristics,
    demonstrating the effect of the relaxation rates $\gamma_{30}$ and
    $\gamma_{31}$ from the vibronic state $|3 \rangle$ on the occurrence
    of bistability.
    The set of parameters used in the simulations are: $\sigma = 0.1 J$,
    $\Delta_{10}^{(0)} = \epsilon_0 - \omega_i - 2.02 J$ (tuning to
    the J-band maximum), $\Gamma = 500\gamma_0$, and $\Gamma_R = 10 \sigma^*$.}
    \label{Fig: Steady State 2}
\end{figure}

From the physical point of view, the most favorable conditions for
bistability occur in the case of slow relaxation from the vibronic
state $|3 \rangle$, which is populated via a fast energy transfer
from the two-exciton state $|2 \rangle$ (fast exciton-exciton
annihilation). Indeed, under these conditions, all population can be
rapidly transferred to the state $|3 \rangle$, and, accordingly, the
system can be made transparent easier as compared to the case of the
one-exciton model. Clearly, faster relaxation from the state $|3
\rangle$ to the ground state $|0 \rangle$ will deteriorate the
condition for the occurrence of bistability, while slower relaxation
improves the situation. Figure~\ref{Fig: Steady State 2}
demonstrates this.

%Fig. 4

%Fig. 5

\begin{figure}[lht]
\begin{center}
\includegraphics[width = 0.9\columnwidth,scale=1]{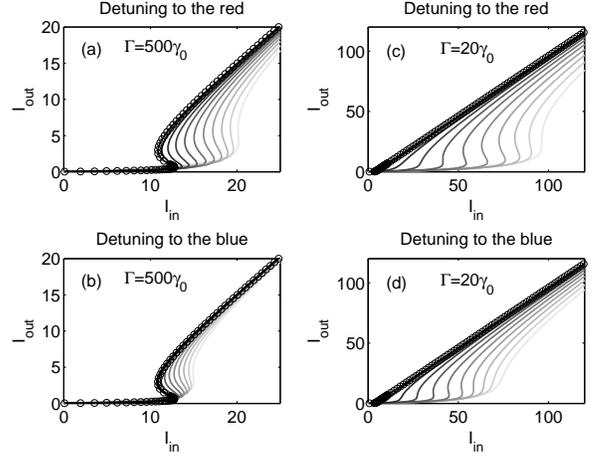}
\end{center}
    \caption{Examples of the output-input characteristics,
    demonstrating
    the combined effect of dephasing, $\Gamma$, and detuning
    off-resonance, $\Delta_{10}$, on the occurrence of bistability.
    In the simulations, the following set of parameters were used:
    a disorder strength $\sigma = 0.1 J$ \,(HWHM $\sigma^* = 0.024 J$),
    the exciton-exciton annihilation rate $w = 5000 \gamma_0$, the
    decay rates of the intermediate vibronic level $\gamma_{31}=
    \gamma_{30} = \gamma_0$, and the superradiant constant $\Gamma_R =
    10\sigma^*$. Panels (a) and (b) represent the results obtained for $\Gamma =
    500\gamma_0 \sim \sigma^*$ when changing $\Delta_{10}$ from the
    absorption maximum at $\Delta_{10}^{(0)} = \epsilon_0 - \omega_1 -2.02 J$
    to the red (a) and to the blue (b) in 20 steps of $0.0025 J$.
    The lighter curves correspond to a larger $\Delta_{10}$.
    Panels (c) and (d) show similar results obtained for $\Gamma =
    20\gamma_0 \ll \sigma^*$.}
\label{Fig: Steady State}
\end{figure}

\subsection{Effects of detuning}
    \label{Sec: Effect of detuning}

As we mentioned in Sec.~\ref{Sec: Bistability equation}, a naive
viewpoint is that tuning of the incoming field to the absorption
maximum is expected to give the lowest threshold for bistability. In
this section, we show that in general this expectation is incorrect.

In Fig.~\ref{Fig: Steady State} we plotted the results of our
simulations of the film's optical response as a function of the
detuning off-resonance, $\Delta_{10}$, obtained for two values of
the dephasing constant $\Gamma$. The disorder strength was set to
$\sigma = 0.1 J$, resulting in an inhomogeneous HWHM $\sigma^* =
0.024 J$. From these data, one can distinguish two regimes. First,
for a relatively large $\Gamma = 500\gamma_0 = 0.02 J \sim \sigma^*$
[panels (a) and (b)] the film's response behaves according to the
naive reasoning: the output-input characteristic looses its S-shaped
form upon a deviation of the incoming field frequency from the
absorption maximum. In contrast, as is observed in Figs.~\ref{Fig:
Steady State}(c) and (d), for $\Gamma = 20\gamma_0\ll \sigma^*$,
when the absorption width is dominated by inhomogeneous broadening
$\sigma^*$, tuning away from the resonance favors bistability.

We note that similar behavior has been found for assemblies of
inhomogeneously broadened two-level emitters placed in a
cavity,~\cite{Gibbs76,Bonifacio78,Hassan78} where it was suggested
that this counterintuitive frequency dependence results from the
interplay of absorptive and dispersive contributions to the
nonlinear susceptibility. We believe that our model exhibits the
same spectral behavior because only the ground state to one-exciton
transitions lead to spectral sensitivity. The one-to-two exciton
transitions and the relaxation from the molecular vibronic level do
not: the former because of the fast exciton annihilation, which
washes out all spectral details, and the latter because it occurs
from a relaxed state. Thus, all spectral features of the two-exciton
model of the film's bistability are driven by the ground state to
one-exciton transitions. In other words, the one-exciton (two-level)
model considered in our previous paper~\cite{Klugkist07a} is
relevant for explaining the observed spectral behavior. In this
case, the bistability equation~(\ref{Eq: Steady state output field})
is reduced to
%\begin{widetext}
\begin{align}
    \label{Eq: Steady state output field two-level model}
   |\Omega_i|^2
   = \Bigg\{\left[1 + \gamma_R \frac{N_s}{N} \bigg\langle \mu_{10}^2
   \frac{\Gamma_{10}}{\Gamma_{10}^2 + \Delta_{10}^2
   + |\Omega|^2 \Gamma_{10}/\gamma_0} \bigg\rangle \right]^2
   \nonumber\\
   \nonumber\\
   + \left[ \gamma_R \frac{N_s}{N}
   \bigg\langle \mu_{10}^2 \frac{\Delta_{10}}{\Gamma_{10}^2
   + \Delta_{10}^2 + |\Omega|^2 \Gamma_{10}/\gamma_0}
   \bigg \rangle \right]^2 \Bigg\} |\Omega|^2 \ .
\end{align}
%\end{widetext}

In our further analysis we show that, indeed, the interplay of the
absorptive and dispersive terms in Eq.~(\ref{Eq: Steady state output
field two-level model}) is responsible for the counterintuitive
spectral behavior. First, let us assume that we are far outside the
resonance, i.e., $|\Delta_{10}|$ is large compared to the absorption
HWHM, whether the homogeneous ($\Gamma^* = \langle \Gamma_{10}
\rangle$) or the inhomogeneous one ($\sigma^*$). Then, the
dispersive term drives the bistability, because its magnitude
decreases as $|\Delta_{10}|^{-1}$ upon increasing $\Delta_{10}$,
while the absorptive one drops faster, proportionally to
$\Delta_{10}^{-2}$. The critical superradiant constant for the
dispersive bistability has been reported to be $\Gamma_R^c =
4\big[\Gamma^* + (\Gamma^{*2} + \Delta_{10}^2)^{1/2} \big]$ (see,
e.g., Ref.~\onlinecite{Gibbs76}) which is reduced to $\Gamma_R^c
\approx 4|\Delta_{10}|$ in the limit of $|\Delta_{10}| \gg
\Gamma^*$. On the other hand, we found within the one-exciton
model~\cite{Klugkist07a} that close to the resonance ($|\Delta_{10}|
\ll \sigma^*$), where the contribution of the absorptive term is
dominant, $\Gamma_R^c$ scales superlinearly with the HWHM, namely as
$(\sigma^*/\Gamma^*)^{\alpha}\Gamma^*$ with $\alpha \approx 1.7$.
Similar scaling ($\Gamma_R^c = \sigma^{*2}/\Gamma^*$) has been
obtained in Ref.~\onlinecite{Hassan78} for a collection of
inhomogeneously broadened two-level systems placed in a cavity.

The superlinear dependence of $\Gamma_R^c$ for the absorptive type
of bistability is a key ingredient in understanding the
counterintuitive $\Delta_{10}$ behavior of the film's optical
response. Indeed, let $|\Delta_{10}| \gg \sigma^*$ and $\Gamma_R =
4|\Delta_{10}|$, i.e., we are at the (dispersive) bistability
threshold. Now, let us go back to the resonance, where bistability
is of absorptive nature. Choose for the sake of simplicity
$\Gamma_R^c = \sigma^{*2}/\Gamma^*$ as the critical value. If
$4|\Delta_{10}| > \sigma^{*2}/\Gamma^*$, we are still above the
(absorptive) bistability threshold, while in the opposite case
bistable behavior is not possible. For $\sigma^* \sim \Gamma^*$, the
line width is almost of homogeneous nature, and tuning away from the
resonance deteriorates the conditions for the occurrence of
bistability.~\cite{Gibbs76} In our simulations, this holds for the
case of $\Gamma = 500 \gamma_0 = 0.02 J$ and $\sigma^* = 0. 024 J$
[see panels (a) and (b) in Fig.~\ref{Fig: Steady State}].

To conclude this section, we note that the detuning effect found in
our simulations is asymmetric with respect to the sign of
$\Delta_{10}$: the behavior of $I_{\mathrm{out}}$ versus
$I_{\mathrm{in}}$ is different for the incoming frequency tuned to
the red or to the blue from the absorption maximum. We believe that
this arises from the asymmetry of the absorption spectrum.

\subsection{Phase diagram}
    \label{Sec: Phase diagram}

%Fig. 6

\begin{figure}[lht]
\begin{center}
\includegraphics[width = 0.9\columnwidth]{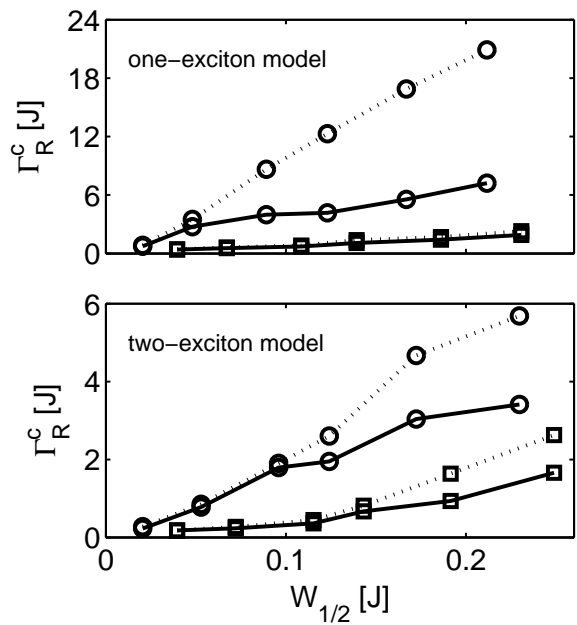}
\end{center}
    \caption{Phase diagram of the bistable optical response of
    a thin film in the ($\Gamma_R,W_{1/2}$)-space, where $W_{1/2} = \sigma^*
    + \Gamma^*$, with $\Gamma^* = \Gamma + \langle \gamma_{10}
    \rangle/2$,
    is used as a measure for HWHM of the absorption spectrum accounting for
    contributions of inhomogeneous and homogeneous broadening (through
    $\sigma^*$ and $\Gamma^*$, respectively) to the total width
    of the $J$ band. The data were obtained by solving
    Eq.~(\protect\ref{Eq: Steady state output field}) for the input
     field tuned to the J-band center, $\Delta_{10}^{(0)} = \epsilon_0 -
    \omega_i - 2.02 J$ and varying the disorder strength $\sigma$. In
    both panels, upper and lower curves correspond to $\Gamma = 20 \gamma_0$
    and $\Gamma = 500 \gamma_0$, respectively.
    The open circles and squares represent the numerical data points,
    whereas the solid lines are a guide to the eye. The solid lines
    themselves represent the $W_{1/2}$-dependence of the critical
    superradiant constant $\Gamma_R^c$. Above (below) the curve for a
    given $\Gamma$, the film behaves in a bistable (stable)
    fashion. For comparison, we also plotted the phase diagram calculated
    under the assumption that the detuning is the only stochastic parameter
    (dotted curves, cf. Ref.~\protect\onlinecite{Glaeske01}).}
\label{Fig: Phase diagram sigma}
\end{figure}

In Fig.~\ref{Fig: Phase diagram sigma} we plotted the results of a
comparative study of phase diagrams of the film's response
calculated within the one- and two-exciton model under the resonance
condition, $\Delta_{10}^{(0)} = \epsilon_0 - \omega_i - 2.02 J$.
Presented is the critical superradiant constant $\Gamma_R^c$ versus
the quantity $W_{1/2} = \sigma^* + \Gamma^*$, where $\Gamma^* =
\Gamma + \langle \gamma_{10} \rangle/2$ is the homogeneous width of
the one-exciton transition. The last term denotes the averaged rate
of population relaxation from the one-exciton state to the ground
state, see Eq.~(\ref{Eq: rho11}). Roughly, $W_{1/2}$ can be
interpreted as the HWHM of the absorption spectrum accounting for
both inhomogeneous and homogeneous broadening (through $\sigma^*$
and $\Gamma^*$, respectively). The upper (lower) solid curve in both
panels was obtained for the dephasing constant $\Gamma = 20\gamma_0$
($\Gamma = 500\gamma_0$) and varying the disorder strength $\sigma$.
For a given $\Gamma$, the film is bistable (stable) above (below)
the corresponding curve. To compare these results with those
calculated under the assumption that the detuning is the only
stochastic parameter,~\cite{Glaeske01} we also plotted the
$\Gamma_{\mathrm{R}}^c$ vs $W_{1/2}$ dependence taking all the
transition dipole moments and relaxation constants equal to their
averaged values (dotted curves).

One of the principal conclusions which can be drawn from the data in
Fig.~\ref{Fig: Phase diagram sigma} is that a more efficient
dephasing helps the occurrence of bistability: all curves calculated
for $\Gamma = 20\gamma_0$ lie above those obtained for $\Gamma =
500\gamma_0$. The physics of this behavior is simple: as the
threshold for the absorptive bistability is $\Gamma_R^c =
(\sigma/\Gamma^*)^{\alpha}\Gamma^*$ (see Sec.~\ref{Sec: Effect of
detuning}), a smaller $\Gamma^*$ gives rise to a higher threshold
value for $\Gamma_R$. Thus, adjusting the dephasing constant
$\Gamma^*$, we can manipulate the film's optical response. This
conclusion has been drawn already in Ref.~\onlinecite{Malyshev00}
within the simplified one-exciton model.

Another observation is that the magnitude of the critical
superradiant constant $\Gamma_R^c$ is considerably lower in the
two-exciton model than in the one-exciton approach. This was to be
expected from the physical reasoning which we presented above: a
fast exciton-exciton annihilation combined with a slow relaxation
from the high-lying molecular vibronic level favors bistability.
Without showing detailed data, we note that also the critical
switching intensity, i.e., the intensity calculated at the
bistability threshold, is smaller in the two-exciton model compared
to the one-exciton model. In both models, it also decreases upon
increasing the dephasing rate.

Finally, from comparison between the solid and dotted curves in
Fig.~\ref{Fig: Phase diagram sigma}, it appears that, surprisingly,
bistability is favored by the fact that also transition dipole
moments and relaxation constants are stochastic variables and not
only the detuning, as was assumed in the simplified model of
Ref.~\onlinecite{Glaeske01}. At first glance this seems
counterintuitive. However, inspection of changes in the absorption
spectrum allows to shed light of on this result. We found that upon
neglecting the fluctuations, the absorption spectrum, first,
acquires a shift which introduces an additional detuning
off-resonance. Second, the shape of the absorption spectrum gets
more asymmetric. As the film's response is sensitive to both the
detuning and asymmetry, the combined effect of these changes
produces the observed big difference between the two sets of
calculations. In principle, this discrepancy may be reduced by
adjusting the detuning; it is impossible, however, to correct for
asymmetry. Most importantly, this comparison shows that to
adequately calculate the film's optical response, fluctuations of
all variables should be taken into account.

\section{Thin film of PIC: Estimates}
    \label{Sec: Estimates}

In this section we will analyze low-temperature experimental data of
$J$-aggregates of pseudo-isocyanine (PIC) to shed light on the
feasibility of measuring optical bistability in a thin film of PIC.
We will focus, in particular, on aggregates of PIC-Br studied
experimentally in detail in Refs.~\onlinecite{Fidder90,Fidder93}
and~\onlinecite{Minoshima94}. At low temperatures, the absorption
spectrum of PIC-Br is dominated by a very narrow absorption band
(HWHM = 17 cm$^{-1}$) peaked at $\lambda$ = 573 nm and red shifted
relative to the main monomer feature ($\lambda$ = 523 nm). For these
aggregates, vibration-induced intra-band relaxation is strongly
suppressed (no visible Stokes shift of the fluorescence spectrum
with respect to the $J$-band is observed). This favors a long
exciton lifetime, which is highly desirable from the viewpoint of
saturation, and thus for optical bistability. The lifetime of the
exciton states forming the $J$-band in PIC-Br is conventionally
assumed to be of radiative nature. For temperatures below about 40
K, it has been measured to be 70 ps.~\cite{Fidder90}

Within the one-exciton model studied in our previous
paper,~\cite{Klugkist07a} we found that the number density of
monomers, required for the driving parameter $\Gamma_R/\sigma^*$ to
exceed the bistability threshold, has to obey $n_0 > 10^{19}$
cm$^{-3}$. Such densities can be achieved in thin films prepared by
the spin-coating method.~\cite{Misawa93,Ramunas} Within the extended
four-level model considered in the present paper, the critical ratio
of $\Gamma_R/\sigma^*$ may be even lower. Thus, we believe that from
the viewpoint of monomer density, J-aggregates of PIC are promising
candidates.

Another important requirement for candidates, potentially suitable
for bistable devices, is their photostability. J-aggregates are
known to bleach if they are exposed for a long time to powerful
irradiation. Therefore, it is useful to estimate the electromagnetic
energy flux through the film. For the field slightly below the
higher switching threshold, the dimensionless intensity inside the
film obeys $I_{\mathrm{out}} = |\Omega|^2/(\gamma_0\sigma^*) \approx
1$ (see, e.g., Fig.\ref{Fig: Steady State GammaR}). Using the
expression for the monomer spontaneous emission rate $\gamma_0 =
32\pi^3d_0^2/(3\hbar\lambda^3)$, we obtain $E_{\mathrm{out}}^2
\approx 32 \pi^3\hbar\sigma^*/(3\lambda^3)$. The electromagnetic
energy flux through the film is determined by the Poynting vector,
whose magnitude is given by $S_{\mathrm{out}} =
cE^2_{\mathrm{out}}/(4\pi)$. Being expressed in the number of
photons $S_{\mathrm{out}}/(\hbar\omega_{10})$, passing per cm$^2$
and per second through the film, this value corresponds to $5 \times
10^{21}$ photons/(cm$^{2}$\,s). As is seen from Fig.~\ref{Fig:
Steady State GammaR}, above the switching threshold the intensity
inside the film rises by an order of magnitude. Hence, above
threshold the electromagnetic energy flux reaches a value on the
order of $S_{\mathrm{out}} \approx 5\times 10^{22}$
photons/(cm$^{2}$\,s).

Furthermore, the typical time $\tau$ for the outgoing intensity
$I_{\mathrm{out}}$ to reach its stationary value is on the order of
the population relaxation time, which is 70 ps, except for values of
$I_{\mathrm{out}}$ slightly above (below) the higher (lower)
switching threshold, where the relaxation slows
down.~\cite{Klugkist07a} This means that typically, a nanosecond
pulse is enough to achieve the steady-state regime. Bearing in mind
the above estimates for $S_{\mathrm{out}}$, we obtain the
corresponding flux for a nanosecond pulse $S_{\mathrm{out}} \approx
10^{13}$ photons/(cm$^2$\, ns). On the other hand, for a thin film
of thickness $L = \lambda/(2\pi)$ and number density of monomers
$n_0 = 10^{20}$ cm$^{-3}$, the surface density is $n_0
\lambda/(2\pi) \approx 10^{15}$ cm$^{-2}$. Combining these numbers,
we conclude that only one photon per $20$ monomers produces the
effect, which is well below the bleaching threshold.~\cite{Ramunas}

\section{Summary and concluding remarks}
    \label{Summary}

We theoretically studied the optical response of an ultrathin film
of oriented J-aggregates with the goal to examine the effect of
two-exciton states and exciton-exciton annihilation  on the
occurrence of bistable behavior. The standard Frenkel exciton model
was used to describe a single aggregate: an open linear chain of
monomers coupled by delocalizing dipole-dipole excitation transfer
interactions, in combination with uncorrelated on-site disorder,
which tends to localize the exciton states.

We considered a single aggregate as a meso-ensemble of exciton
localization segments, ascribing to each segment a four-level system
consisting of the ground state (all monomers in the ground state),
an $s$-like one-exciton state, a two-exciton state constructed as
the antisymmetric combination of this $s$-like state and an
associated $p$-like one-exciton state, and a vibronic state of the
monomer through which the two-exciton states annihilate. To select
the $s$- and $p$-like states, a new procedure was employed which
correctly accounts for the fluctuations and correlations of the
transition energies and transition dipole moments, improving on
earlier works.~\cite{Glaeske01} The optical dynamics of the
localization segment was described within the $4\times4$-density
matrix formalism, coupled to the total electromagnetic field. In the
latter, in addition to the incoming field, we accounted for a part
produced by the aggregate dipoles.

We derived a novel steady-state equation for the transmitted signal
and demonstrated that three-valued solutions to this equation exist
in a certain domain of the multi-parameter space. Analyzing this
equation, we found that several conditions promote the occurrence of
bistable behavior. In particular, a fast exciton-exciton
annihilation, in combination with a slow relaxation from the monomer
vibronic state, favors bistablity. In contrast, fast relaxation from
the vibronic level to the ground state acts against the effect.
Additionally, a faster dephasing also works in favor of the
occurrence of bistability.

The interplay of detuning away from the resonance and dephasing was
found to be counterintuitive. When homogeneous broadening of the
exciton states (associated with the incoherent exciton-phonon
scattering) is comparable with the inhomogeneous broadening
(resulting from the localized nature of the exciton states), the
detuning destroys bistability. Oppositely, at a slower dephasing,
the bistability effect is favored by tuning away from the resonance.
We relate this anomalous behavior to an interplay of the absorptive
and dispersive parts of the nonlinear susceptibility, which jointly
contribute to the overall effect.

We found that in general, including the one-to-two-exciton
transitions promotes bistability. All critical parameters, such as
the critical superradiant constant, driving the bistability, and the
critical switching intensity are lower than in the one-exciton
model.~\cite{Klugkist07a}  In addition, bistable behavior is easier
to reach if the ratio of the inhomogeneous and homogeneous width is
reduced. We also found that the stochastic nature of the transition
dipole moments (the aspect in which our model goes beyond
Ref.~\onlinecite{Glaeske01}) strongly influences the film's optical
response.

Estimates of parameters of our model for aggregates of polymethine
dyes at low temperatures indicates that a film with a monomer number
density on the order of $10^{20}$ cm$^{-1}$ and a thickness of
$\lambda/2\pi$, achievable with the spin coating
method,~\cite{Misawa93} is sufficient to realize the effect. Under
these conditions, one photon per 20 monomers produces the switching
of the film's transmittivity.

To conclude, we point out that a microcavity filled with molecular
aggregates~\cite{Lidzey98,Litinskaya04,Beltyugov04,Agranovich05,Zoubi05}
in the strong coupling regime of excitons to cavity modes is another
promising arrangement to realize an all-optical switch. The recent
observation of optical bistability in planar {\it inorganic}
microcavities~\cite{Baas04} and the prediction of the effect for
hybrid {\it organic-inorganic} microcavities~\cite{Zoubi07} in the
strong coupling regime suggest that {\it organic} microcavities can
exhibit a similar behavior.

\acknowledgments

This work is part of the research program of the Stichting voor
Fundamenteel Onderzoek der Materie (FOM), which is financially
supported by the Nederlandse Organisatie voor Wetenschappelijk
Onderzoek (NWO). Support was also received from NanoNed, a national
nanotechnology programme coordinated by the Dutch Ministry of
Economic Affairs.

\end{document}